
\input harvmac

\def\np#1#2#3{Nucl. Phys. {\bf B#1} (#2) #3}
\def\pl#1#2#3{Phys. Lett. {\bf #1B} (#2) #3}

\def\physrev#1#2#3{Phys. Rev. {\bf D#1} (#2) #3}

\def\prep#1#2#3{Phys. Rep. {\bf #1} (#2) #3}

\def\cmp#1#2#3{Comm. Math. Phys. {\bf #1} (#2) #3}
\def\vev#1{\langle#1\rangle}
\def\Tr{{\rm Tr ~}}

\def\CW{{\cal W}}
\def\tilde{\widetilde}

\Title{hep-th/9411149, RU-94-82, IASSNS-HEP-94/98}
{\vbox{\centerline{Electric-Magnetic Duality in Supersymmetric}
\centerline{Non-Abelian Gauge Theories}}}
\bigskip
\centerline{N. Seiberg}
\vglue .5cm
\centerline{Department of Physics and Astronomy}
\centerline{Rutgers University}
\centerline{Piscataway, NJ 08855-0849, USA}
\vglue .3cm
\centerline{and}
\vglue .3cm
\centerline{School of Natural Sciences}
\centerline{Institute for Advanced Study}
\centerline{Princeton, NJ 08540, USA}

\bigskip\bigskip

\noindent
We demonstrate electric-magnetic duality in N=1 supersymmetric
non-Abelian gauge theories in four dimensions by presenting two
different gauge theories (different gauge groups and quark
representations) leading to the same non-trivial long distance physics.
The quarks and gluons of one theory can be interpreted as solitons
(non-Abelian magnetic monopoles) of the elementary fields of the other
theory.  The weak coupling region of one theory is mapped to a strong
coupling region of the other.  When one of the theories is Higgsed by an
expectation value of a squark, the other theory is confined.  Massless
glueballs, baryons and Abelian magnetic monopoles in the confining
description are the weakly coupled elementary quarks (i.e.\ solitons of
the confined quarks) in the dual Higgs description.

\Date{11/94}

\newsec{Introduction}

\nref\ads{I. Affleck, M. Dine, and N. Seiberg, \np{241}{1984}{493};
\np{256}{1985}{557}}%
\nref\nsvz{V.A. Novikov, M.A. Shifman, A. I.  Vainstain and V. I.
Zakharov, \np{223}{1983}{445}; \np{260}{1985}{157}}%
\nref\rusano{V. Novikov, M. Shifman, A. Vainshtein and V.
Zakharov, \np{229}{1983}{381}}
\nref\svholo{M.A. Shifman and A.I Vainshtein, \np{277}{1986}{456};
\np{359}{1991}{571}}%
\nref\cern{D. Amati, K. Konishi, Y. Meurice, G.C. Rossi and G.
Veneziano, \prep{162}{1988}{169} and references therein}%
\nref\nonren{N. Seiberg, hep-ph/9309335, \pl{318}{1993}{469}}%
\nref\nati{N. Seiberg, hep-th/9402044, \physrev{49}{1994}{6857}}%
\nref\kl{V. Kaplunovsky and J. Louis, hep-th/9402005,
\np{422}{1994}{57}}%
\nref\ils{K. Intriligator, R.G. Leigh and N. Seiberg, hep-th/9403198,
\physrev{50}{1994}{1092}; K. Intriligator, hep-th/9407106,
\pl{336}{1994}{409}}%
\nref\swi{N. Seiberg and E. Witten, hep-th/9407087, \np{426}{1994}{19}}%
\nref\swii{N. Seiberg and E. Witten, hep-th/9408099, RU-94-60,
IASSNS-HEP-94/55}%
\nref\intse{K. Intriligator and N. Seiberg, hep-th/9408155, RU-94-68,
IASSNS-HEP-94/67}%
\nref\iss{K. Intriligator, N. Seiberg and S. Shenker, hep-ph/9410203,
RU-94-75, IASSNS-HEP-94/79}%

The holomorphic objects in supersymmetric field theories in four
dimensions can be analyzed exactly \refs{\ads-\iss} (for a recent short
review see
\ref\pwer{N. Seiberg, The Power of Holomorphy -- Exact Results in 4D
SUSY Field Theories.  To appear in the Proc. of PASCOS 94.
hep-th/9408013, RU-94-64, IASSNS-HEP-94/57.}).
These objects characterize much of the long distance physics of these
theories including their phase structure and their light spectrum (in
some cases even the massive spectrum can be determined).  This fact
makes these theories interesting laboratories for the study of the
behavior of gauge theories, leading to new insights about issues like
confinement and chiral symmetry breaking.

We also hope that these solvable examples will address more fundamental
issues in field theory and string theory.  One such issue, which we will
discuss here, is the weak-strong coupling duality between electric and
magnetic variables
\ref\om{C. Montonen and D. Olive, \pl {72}{1977}{117}; P. Goddard,
J. Nuyts, and D. Olive, Nucl. Phys. {\bf B125} (1977) 1.}.
In N=4
\ref\dualnf{H. Osborn, \pl{83}{1979}{321}; A. Sen, hep-th/9402002,
TIFR-TH-94-03; hep-th/9402032,TIFR-TH-94-08; C. Vafa and E. Witten,
hep-th/9408074, HUTP-94/A017, IASSNS-HEP-94-54}
and certain N=2 \swii\ supersymmetric theories this duality is expected
to be an exact symmetry of the full theory.  These theories are
characterized by scale invariance for every value of the coupling
constant $\tau$ on which the duality transformations act.  In
asymptotically free theories the coupling constant $\tau$ is replaced by
a dynamically generated scale $\Lambda$ and therefore it seems that
duality cannot act.  However, in some N=2 \refs{\swi,\swii} and
N=1 \intse\ theories with an unbroken Abelian gauge group, i.e.\ an
Abelian Coulomb phase, duality in the value of the effective coupling
constant $\tau_{eff}$ of the low energy photons is crucial in the
understanding of the Coulomb phase.

In this paper we will study asymptotically free theories in their
non-Abelian Coulomb phase, i.e. a phase in which the light spectrum
includes non-Abelian gluons and quarks.  We will show how
duality helps to understand this phase.

The prototype of our theories is supersymmetric QCD \refs{\ads,\cern}.
This theory is based on an $SU(N_c)$ gauge theory with $N_f$ flavors of
quarks, $Q^i$ in the $N_c$ representation and $\tilde Q_{\tilde i}$ in
the $\overline N_c$ representation ($i, \tilde i =1, \dots, N_f$).  The
anomaly free global symmetry is
\eqn\globsym{SU(N_f)\times SU(N_f) \times U(1)_B \times U(1)_R }
where the the quarks transform as
\eqn\qtransl{\eqalign{
Q &\qquad (N_f,1,1,{N_f-N_c \over N_f}) \cr
\tilde Q & \qquad (1, \overline N_f,-1,{N_f-N_c \over N_f}). \cr}}
The interesting gauge invariant operators which we will study
are
\eqn\opsin{\eqalign
{&M^i_{\tilde i}=Q^i Q_{\tilde i} \cr
&B^{[i_1,...,i_{N_c}]}=Q^{i_1}...Q^{i_{N_c}} \qquad {\rm for}~ N_f
\ge N_c \cr
&\tilde B_{[\tilde i_1,...,\tilde i_{N_c}]}=\tilde Q_{\tilde
i_1}... Q_{\tilde i_{N_c}} \qquad {\rm for}~ N_f  \ge N_c .\cr}}

For $N_f<N_c$ the quantum theory does not have a ground state \ads.  For
$N_f\ge N_c$ the quantum theory has a moduli space of inequivalent vacua
\refs{\ads,\nati}.  For $N_f=N_c$ this space is different {}from the
classical one \nati.  For $N_f=N_c+1$ the two spaces are identical but
the interpretation of the singularity at the origin is different --
there are massless mesons and baryons \nati.

For $N_f \ge N_c+2$ the quantum moduli space is the same as the
classical one \nati.  This can be shown by turning on a tree level
mass term $W_{tree}=\Tr mM$ and finding
\eqn\mexpmass{\vev{M^i_{\tilde i}} = \Lambda^{3N_c-N_f \over N_c} (\det
m)^{1 \over N_c} \left({1 \over m}\right)^i_{\tilde i} .}
Then, by studying various limits of $m \rightarrow 0$, all the classical
values of $M$ with $B=\tilde B=0$ can be obtained.  In particular, the
point $M=B=\tilde B=0$ with unbroken $SU(N_f)\times SU(N_f) \times
U(1)_B\times U(1)_R$ is on the quantum moduli space.

It was conjectured in \nati\ that the theory at the origin is in a
non-Abelian Coulomb phase.  Clearly, this is the case for $N_f \ge 3N_c$
where the theory is not asymptotically free.  Hence, the IR theory is a
free theory of the elementary quarks and gluons.  The following simple
argument shows that there are massless quarks and gluons at the origin
also for smaller values of $N_f$ -- at least for $N_f \ge N_c +4$.
Consider a classical flat direction along which $SU(N_c)$ is broken to
$SU(k)$.  It is easy to see that classically there are $N_f-N_c+k$
flavors in the low energy theory.  If $N_f-N_c+ k \ge 3k$ the low energy
theory is not asymptotically free.  Therefore, in the full quantum
theory these quarks and gluons will remain massless.  Taking the limit
to the origin of the moduli space, we learn that $SU\left( \left[
{N_f-N_c \over 2} \right] \right)$ is a subgroup of the gauge group
there and the theory is in a non-Abelian Coulomb phase.

This paper is devoted to the study of the non-Abelian Coulomb phase.  We
will argue that for $ 3 N_c/2 < N_f < 3N_c$ the theory at the origin of
the moduli space is an interacting conformal field theory of quarks and
gluons.  This theory has two dual descriptions.  In the original
``electric'' variables it is an $SU(N_c)$ gauge theory with $N_f$
flavors.  The dual theory is based on ``magnetic'' variables.  It is an
$SU(N_f-N_c) $ theory with $N_f$ flavors and an additional gauge
invariant massless field.  The quarks and the gluons of one description
can be thought of as solitons (magnetic monopoles) of the quarks and the
gluons of the dual theory.

As expected with such dual theories, when one of them is in a weakly
coupled Higgs phase its dual is in a strongly coupled confining phase.
For $N_f \ge 3N_c$ the original theory is not asymptotically free.  The
electric variables are free in the infrared and the magnetic ones are
infinitely strongly coupled.  The theory is thus in a ``free non-Abelian
electric phase.'' For $N_c+2 \le N_f \le 3N_c/2 $ the electric variables
are very strongly coupled but the theory of the magnetic variables is
not asymptotically free and hence it is free in the IR.  The theory is
thus in a ``free non-Abelian magnetic phase.''  These two phases are
dual to each other while the interacting non-Abelian Coulomb phase is
self dual.

It is clear that the theory and its dual should have the same global
symmetries.  Global symmetries are associated with observable currents
which must be the same in the two theories.  Gauge symmetries, on the
other hand, are not real symmetries; they are a redundancy in the
description.  Therefore, it is possible, and in fact it is the case,
that the gauge group of the dual theory is in general different {}from
the original one.  However, at a more intuitive level, one might wonder
how we can have two descriptions of the same theory with a different
number of gluons.  The answer is that the duality makes sense in
interacting scale invariant theories.  Such theories do not have a well
defined particle interpretation and therefore they can be described by
different sets of massless interacting particles.

We will show similar phenomena in an $SO(N_c)$ theory with $N_f$
flavors in the $N_c$ representation.  The dual of this theory is an
$SO(N_f-N_c+4)$ gauge theory with $N_f$ flavors and an additional gauge
invariant massless field.  We will not discuss $SP(N_c)$ theories in
detail here, but will simply note that the dual of an $SP(N_c)$ theory
with $2N_f$ quarks in the fundamental representation is an
$SP(N_f-N_c-2)$ gauge theory with $2N_f$ quarks and another gauge
invariant field.

The gauge group does not change under duality for $SU(N_c)$ with
$N_f=2N_c$, for $SO(N_c)$ with $N_f=2(N_c-2)$ and for $SP(N_c)$ with
$N_f=2(N_c+1)$.  Precisely for these values we can add a chiral field
in the adjoint of the gauge group and an appropriate Yukawa coupling
so that the resulting N=2 supersymmetric theory is scale invariant for
every $\tau$ and therefore it is likely to be fully dual.  This
observation could be important in a deeper understanding of the duality
presented here.

Supersymmetric theories are believed to be relevant to the solution
the hierarchy problem.  Hopefully, our results about the dynamics of
these theories will be helpful in constructing models of dynamical
supersymmetry breaking.  As we mention the potential phenomenological
applications of our results, we cannot resist stating the speculation
that perhaps some, or even all, of the known ``elementary particles''
of the standard model are dual to the truly elementary particles at
shorter distance!

In section 2 we discuss supersymmetric QCD for $ 3N_c/2 < N_f < 3N_c$ in
their interacting non-Abelian Coulomb phase.  In section 3 we describe
the equivalent dual description of these theories.  Just as the
original, electric description can be extended to $3N_c \le N_f$ where
the theory is free, the dual, magnetic theory can be extended to $ N_c
+2 \le N_f \le 3N_c/2$ where it is free.  In section 4 we consider
various deformations of the theories, shedding more light on the
relation between duality, confinement and the Higgs mechanism.  In
section 5 we extend our analysis to $SO(N_c)$ theories, where we gain
further insight about the phases and show how our duality is a
generalization of the duality in Abelian theories.

\newsec{Interacting non-Abelian Coulomb phase for $SU(N_c)$ with
$3N_c/2 < N_f < 3N_c$}

By the general rules of quantum field theory the low energy theory
must be scale invariant.  If it includes interacting quarks and
gluons, the beta function must have a non-trivial zero.  The exact
beta function in supersymmetric QCD satisfies \refs{\rusano,\svholo}
\eqn\betafun{\eqalign{
\beta(g)&= - {g^3 \over 16 \pi^2} {3N_c - N_f + N_f \gamma(g^2) \over 1
- N_c {g^2 \over 8 \pi^2}} \cr
\gamma(g^2) & = -{g^2 \over 8 \pi^2}{N_c^2 -1 \over N_c} + \CO( g^4),
\cr}}
where $\gamma(g^2)$ is the anomalous dimension of the mass.  Since there
are values of $N_f$ and $N_c$ where the one loop beta function is
negative but the two loop contribution is positive, there might be a
non-trivial fixed  point
\ref\bankszaks{T. Banks and A. Zaks, \np{196}{1982}{189}}.
Indeed, by taking $N_c$ and $N_f$ to infinity holding $N_c g^2$ and
${N_f \over N_c}=3 - \epsilon$ fixed, one can establish the existence of
a zero of the beta function at $N_cg_*^2={8 \pi^2 \over 3} \epsilon +
\CO(\epsilon^2) $.  Therefore, at least for large $N_c$ and $\epsilon=
3-{N_f \over N_c} \ll 1$ there is a non-trivial fixed point.  We claim
that such a fixed point exists for every $3N_c/2 < N_f < 3N_c$.

Given that such a fixed point exists, we can use the superconformal
algebra to derive some exact results about the theory.  This algebra
includes an $R$ symmetry.  It follows {}from the algebra that the
dimensions of the operators satisfy
\eqn\dimchir{D \ge {3 \over 2} |R|;}
the inequality is saturated for chiral operators, for which $D = {3
\over 2} R $, and for anti-chiral operators, for which $D =- {3 \over 2}
R $.  Exactly as in N=2 theories in two dimensions this fact leads to
important consequences.  Consider the operator product of two chiral
operators $\CO_1(x) \CO_2(0)$.  All the operators in the resulting
expansion have $R=R(\CO_1) +R(\CO_2)$ and hence $D\ge
D(\CO_1)+D(\CO_2)$.  Therefore, there is no singularity in the expansion
at $x=0$ and we can define the product of the two operators by simply
taking the limit of $x$ to zero.  If this limit does not vanish, it
leads to a new chiral operator $\CO_3$ whose dimension is
$D(\CO_3)=D(\CO_1)+D(\CO_2)$.  We conclude that the chiral operators
form a ring.

Clearly, this $R$ symmetry cannot be anomalous and should commute with
the flavor $SU(N_f) \times SU(N_f) $ symmetry.  Therefore, this $R$
symmetry must be the one appearing in \globsym\qtransl.  Hence the gauge
invariant operators $\tilde Q Q$ have \pwer
\eqn\rdqtq{D(\tilde Q Q)= {3 \over 2} R(\tilde Q Q) = 3 {N_f-N_c \over
N_f}}
and similarly
\eqn\brdqtq{D(B)=D (\tilde B)= {3N_c(N_f-N_c) \over 2N_f}.}
The value of $D(\tilde Q Q)$ was determined in \nati\ {}from \betafun\
-- at the zero of the beta function $\gamma=-3{N_c \over N_f} +1$ and
hence $D=\gamma+2= 3 {N_f-N_c \over N_f}$.

The complete list of unitary representations of the superconformal
algebra was given in
\ref\supconfra{M. Flato and C. Fronsdal, Lett. Math. Phys. {\bf 8}
(1984) 159; V.K. Dobrev and V.B. Petkova, \pl{162}{1985}{127}}
by extending the analysis of the ordinary conformal algebra of ref.
\ref\mack{G. Mack, \cmp{55}{1977}{1}}.
Clearly, all the gauge invariant operators in the theory should be in
unitary representations.  One of the constraints on the representations
which follows already {}from the analysis of \mack\ is that spinless
operators have $D \ge 1$ (except the identity operator with $D=0$) and
the bound is saturated for free fields (satisfying $\partial_\mu
\partial^\mu \Phi =0$).  For $D<1$ ($D\not= 0$) a highest weight
representation includes a negative norm state which cannot exist in a
unitary theory.

The value of $D(\tilde Q Q)$ in \rdqtq\ is inconsistent with this bound
for $N_f < 3N_c/2$.  Therefore, for these values of $N_f$ and $N_c$ our
description of the physics cannot be correct.  We suggest that the
description in this section is valid in the entire range $3N_c/2< N_f <
3N_c$.  For smaller values of $N_f$ a totally different description
should be used.  The fact that the dimension of $M=\tilde Q Q$ becomes
one for $N_f=3N_c/2$ shows that $M$ becomes a free field, i.e.
$\partial^2 M=0$.  This indicates that in the correct description for
$N_f=3N_c/2$ the field $M$, and perhaps even the whole IR theory, is
free.

We cannot prove our assumption that for every $3N_c/2 < N_f < 3N_c$ the
theory at the origin is in a non-Abelian Coulomb phase as we described.
However, the consistency of the picture which emerges strongly suggests
that this is indeed the case.

\newsec{The dual theory}

As $N_f$ becomes smaller with fixed $N_c$, the interactions in the
conformal field theory become stronger.  In such a strongly interacting
massless theory it does not make sense to talk about the spectrum of
massless particles.  In fact, we will suggest that the same theories
have an equivalent description in terms of a different set of degrees of
freedom -- dual variables -- which become weakly coupled as $N_f$
becomes smaller.  Unlike the original variables, the dual ones can also
be used for $N_c+2 \le N_f \le 3N_c/2$ where they become free.

Since for $N_f \ge N_c+1$ the moduli space is not modified quantum
mechanically, there is a vacuum at the origin: $M=B=\tilde B=0$ \nati.
At that point the full $SU(N_f)\times SU(N_f) \times U(1)_B \times
U(1)_R $ flavor symmetry is unbroken.  Therefore, there should be $N_f$
flavors of quarks there.  If the gauge group were $SU(N_c)$ there, we
would be faced with the above mentioned problem with the dimensions.  We
will argue that the gauge group there is actually $SU(N_f-N_c)$.
According to the discussion in the previous section such a theory has an
IR fixed point for $3N_c/2 < N_f < 3N_c$ and becomes free for $N_f \le
3N_c/2 $.  Furthermore, all spinless operators have $D \ge 1$.

If these quarks were the elementary ones, the operators $B$ and
$\tilde B$ could not have been constructed.  Instead we will assume
that the quarks are not the elementary ones and will denote them by
$q$ and $\tilde q$. $q$ transforms as $N_f-N_c$ of the color group and
$\tilde q$ transforms as $\overline{N_f-N_c}$.  Their global quantum
numbers under $SU(N_f)\times SU(N_f) \times U(1)_B \times U(1)_R $ are
determined such that $B$ is proportional to $q^{N_f-N_c}$ and $\tilde
B$ is proportional to $ \tilde q^{N_f-N_c}$.  This leads to
\eqn\newquaqu{\eqalign{
q \qquad {\rm in} \qquad &(\overline N_f,1,{N_c \over N_f-N_c},
{N_c\over N_f}) \cr
\tilde q \qquad {\rm in} \qquad &(1, N_f,- {N_c \over N_f-N_c},
{N_c\over N_f}) . \cr }}
Note that the new quarks $q$ transform differently than $Q$ under
$SU(N_f) \times SU(N_f)$.  Also, the baryon number of $q$ is a fraction
and therefore $q$ cannot be represented as a polynomial in the old
quarks $Q$.  Finally, the assignment of the $R$ charge, which was
motivated by the construction of the baryon operators, is anomaly free.

Although we can construct the baryons, we cannot construct the meson
fields $M$.  Therefore, we suggest that they appear as independent
fields
\eqn\mquantumn{M \qquad {\rm in} \qquad (N_f,\overline N_f, 0,
2({N_f-N_c\over N_f})).}

Our quantum numbers allow us to write a superpotential
\eqn\effsup{W= M^i_{\tilde i} q_i \tilde q^{\tilde i}}
(where the color indices are summed).  Such a superpotential is
necessary to give the fields $q$ and $\tilde q$ a mass away {}from the
origin, as will be discussed in detail below.  Therefore, we assume that
the superpotential \effsup\ is present.  Superficially, the low energy
theory has a new meson operator $ q_i \tilde q^{\tilde i}$ which is not
present in the original theory.  However, because of the term \effsup,
this operator is redundant -- its coefficient can be absorbed in a shift
of $M$.

To summarize, we suggest that the new, ``dual'' variables are the mesons
$M$, the $SU(N_f-N_c)$ gluons and the quarks $q$ and $\tilde q$, and
their Lagrangian includes the superpotential \effsup.

This ansatz for the dual variables at the origin satisfies a highly
non-trivial consistency condition.  The 'tHooft anomaly matching
conditions for $SU(N_f)\times SU(N_f) \times U(1)_B \times U(1)_R $ are
satisfied.  We find both in the original $SU(N_c)$ gauge theory and in
the dual $SU(N_f-N_c)$ gauge theory
\eqn\thooftsu{\eqalign{
SU(N_f)^3 \qquad &N_c d^{(3)}(N_f) \cr
SU(N_f)^2U(1)_R \qquad & -{N_c^2 \over N_f}d^{(2)}(N_f) \cr
SU(N_f)^2U(1)_B \qquad & N_cd^{(2)}(N_f) \cr
U(1)_R \qquad &-N_c^2-1 \cr
U(1)_R^3 \qquad &N_c^2-1-2{N_c^4\over N_f^2} \cr
U(1)_B^2U(1)_R \qquad &-2N_c^2 . \cr}}

For $3N_c/2 < N_f < 3 N_c$ the low energy $SU(N_f-N_c)$ gauge theory
with $N_f$ quarks is asymptotically free.  It flows to the non-trivial
IR fixed point discussed in section 2.  The dimensions of the chiral
operators are determined by their $R$ charge.  In particular, we again
get $D(M)= 3{N_f-N_c \over N_f}$.  Note that the superpotential
\effsup\ which has $R=2$ is formally (because strictly speaking it is
a redundant operator) marginal at the fixed point.

As $N_f$ becomes smaller the original theory becomes more strongly
coupled.  Its dual becomes more weakly coupled.  This is a
characteristic behavior of two dual theories.  Therefore, we suggest
that the two theories are dual in the sense of electric-magnetic
duality.  We will refer to the original variables as electric and to the
dual variables as magnetic.  The quarks $q$ and $\tilde q$ and the
$SU(N_f-N_c)$ gluons are interpreted as solitons (non-Abelian magnetic
monopoles) of the elementary fields.  Below we will substantiate this
interpretation by studying various aspects of the two dual theories.

For $N_c+2 \le N_f \le 3N_c/2$ the magnetic theory is not asymptotically
free and the extreme low energy theory is free.  Then the anomaly is not
important and the $R$ symmetry in the superconformal algebra is the
naive one, giving $D(M)=1$.  Since the IR theory is free, there cannot be
two different descriptions.  In fact, only the free magnetic description
makes sense.  We can describe this phase as a free non-Abelian magnetic
phase.  It is dual to the free non-Abelian electric phase which exists
for $N_f \ge 3N_c$.

The interpretation as a dual theory requires that the dual of the dual
is the original theory.  Starting with $SU(N_c)$ with $N_f$ flavors, its
dual is $SU(N_f-N_c)$ with $N_f$ flavors plus the field $M$.  Dualizing
again, we find an $SU(N_c)$ theory with $N_f$ quarks $Q$ and $\tilde Q$
as in our starting point but now with two elementary meson fields $M$
and $N$ with the superpotential
\eqn\dualdual{(M^i_{\tilde i} - Q^i \tilde Q_{\tilde i})
N^{\tilde i}_i.}
Therefore, the fields $M$ and $N$ are massive -- integrating out
$N^{\tilde i}_i$ we find that $M$ satisfies its definition as a meson
field.  Note the change in sign of the second term which is motivated
by ensuring that the dual of the dual is the original theory.  Such a
change in sign is common in duality (and Legendre) transformations.

In the electric variables $M$ has dimension two at the UV fixed point
and acquires the dimension \rdqtq\ at the IR fixed point.  On the other
hand, at the UV fixed point of the magnetic theory it is more natural to
assign dimension one to $M$ and to the quarks.  This change in the
dimensions can be achieved by a multiplicative power of $\Lambda$.  In
what follows we will arrange the powers of $\Lambda$ according to the
dimensions at the UV fixed point of the electric theory.

We conclude that in the range $3N_c/2 < N_f <3N_c$ the theory at the
origin is in an interacting non-Abelian Coulomb phase.  In that phase
there are two dual descriptions of the physics.  The electric one is
more weakly coupled and hence more natural for $2N_c \le N_f <3N_c$ and
the magnetic description is more natural for $3N_c/2 < N_f \le 2N_c$.
The electric description can be continued to the range $3N_c \le N_f$
where it is free.  The magnetic theory is infinitely coupled there.  The
magnetic description can be extended to $N_c+2 \le N_f \le 3N_c/2$ where
it is free and the electric one is infinitely coupled.  This infinite
coupling and the necessity for changing variables to the magnetic
variables in this range can be signaled by the pole in the beta function
\betafun.  If $\gamma(g^2={8\pi^2 \over N_c}) \ge 1-3{N_c \over N_f}$,
the theory hits the pole before the fixed point at the zero of the
numerator.

\newsec{Deforming dual theories}

\subsec{Deforming the original electric theory}

\noindent
{\it Flat directions in the electric theory}

For $N_f \ge N_c+2$ the topology of the quantum moduli space is the same
as that of the classical moduli space.  The latter can be described by
the D-flat directions in the classical Lagrangian.  Up to gauge and global
rotations they are of the form
\eqn\flatdiro{Q=\pmatrix { a_1& & & \cr
&a_2& & \cr
& & .  & \cr
& & & a_{N_c}\cr
& & & \cr
& & & \cr} ; \quad
\tilde Q= \pmatrix {
\tilde a_1& &       &         \cr
   &\tilde a_2&     &         \cr
   &   & .   &         \cr
   &   &     & \tilde a_{N_c}\cr
   &   &     &         \cr
   &   &     &         \cr} }
with
\eqn\flatdircon{|a_i|^2 - |\tilde a_i|^2= {\rm independent ~ of} ~ i }
The gauge invariant description of the space is given in terms of the
constrained observables $M$, $B$ and $\tilde B$ \nati.  Up to global
symmetry transformations they are
\eqn\flatmbtb{\eqalign{
M&=\pmatrix { a_1 \tilde a_1& & &  & & &\cr
&a_2 \tilde a_2 & &  & & &\cr
& & .  & & & & \cr
& & & a_{N_c}\tilde a_{N_c} & & &\cr
& & & & & & \cr
& & & & & & \cr} \cr
B^{1,...,N_c}&=a_1a_2...a_{N_c} \cr
\tilde B_{1,...,N_c}&=\tilde a_1 \tilde  a_2...\tilde a_{N_c} \cr}}
with all other components of $M$, $B$ and $\tilde B$ vanishing.  We
conclude that the rank of $M$ is at most $N_c$.  If it is less than
$N_c$, either $B=0$ with $\tilde B$ having at most rank one or $\tilde
B=0$ with $B$ having at most rank one.  If the rank of $M$ is equal to
$N_c$, both $B$ and $\tilde B$ have rank one and the product of their
eigenvalues is the same as the product of non-zero eigenvalues of $M$.

The physical interpretation of the flat directions is that the gauge
group is Higgsed.  If $B=\tilde B=0$ and $M$ has rank $k$, $SU(N_c)$ is
broken to $SU(N_c-k)$ with $N_f-k$ flavors.

\vskip 2cm
\noindent
{\it Mass terms in the electric theory}

By turning on a mass term and making it large we can reduce the number
of flavors and connect theories with different values of $N_f$.
Clearly, this makes the low energy theory more strongly coupled.  As the
number of flavors is reduced below $3N_c/2$, the low energy theory is
expected to stay in the non-Abelian Coulomb phase but be describable in
the IR only in terms of the dual variables.  Eventually, when the number
of flavors is below $N_c+2$, the theory confines.

For $3N_c/2 < N_f < 3N_c$ the theory is at a non-trivial fixed point.
We can turn on masses for all the quarks and let the theory flow to a
fully massive theory.  Equation \mexpmass\ and the closely related
gluino condensate
\eqn\gluicond{\vev{\lambda \lambda} = \Lambda^{3N_c-N_f \over N_c} (
\det m)^{1 \over N_c}  }
can then be interpreted in two different ways.  First, when all the
masses are much larger than $\Lambda$ the massive quarks can be
integrated out at one loop leading to a low energy theory with a scale
$\Lambda_L^{3N_c}= \det m \Lambda^{3N_c-N_f}$ with gluino condensate
$\vev{\lambda \lambda} = \Lambda_L^{3N_c}$.  Alternatively, when the
masses are much smaller than $\Lambda$, equations \mexpmass\ and
\gluicond\ reflect the anomalous dimensions of the mass operator at
the fixed point.  Since the answer in terms of the mass parameter
$m_*$ at the fixed point should be independent of $\Lambda$, we define
$m_*=m \Lambda^{3N_c-N_f \over N_f}$.  Hence the dimension of $m_*$ is
$3N_c/N_f$ and correspondingly $D(\tilde Q Q)= 3(N_f-N_c)/N_f$ exactly
as in \rdqtq.

\subsec{Deforming the dual, magnetic theory}

\noindent
{\it Flat directions in the magnetic theory}

Superficially the magnetic theory has more flat directions than the
electric theory.  For example, an arbitrary $M$ with $q=\tilde q=0$
looks like a flat direction.  This naive analysis is incorrect.

When $M$ is non-zero it leads to a mass term for some $q$ and $\tilde q$
by the superpotential \effsup.  If $M$ has rank $r$, $r$ flavors of dual
quarks acquire a mass and the remaining $N_f-r$ quarks remain massless.
As long as $r \le N_c-2$ the theory of the massless $q$'s remains in a
non-Abelian Coulomb phase.  Then, the $F$ and $D$ terms in the dual
theory allow us to turn on expectation values either for $q$ or for
$\tilde q$ (but not for both) with equal eigenvalues, thus completely
breaking the $SU(N_f-N_c)$ gauge group.  This corresponds to $B\not= 0$,
with rank one and $\tilde B=0$ or $\tilde B\not= 0$, with rank one and
$B=0$.

If $r \ge N_c-1$, more $q$'s are lifted and the theory is no longer in
a non-Abelian Coulomb phase.  Then, a more careful analysis is
necessary.  We use the flavor symmetry to bring $M$ to the form
\eqn\formm{M=\pmatrix{\hat M & 0 \cr 0& M_0\cr}}
with $M_0$ a square matrix with $N_c-1$ rows and rank $N_c-1$.  For
simplicity, we consider a flat direction where the eigenvalues of
$M_0$ are much larger than the entries in the $N_f-N_c+1 $ dimensional
matrix $\hat M$.  We first integrate out the heavy $q$'s to find a low
energy theory with gauge group $SU(N_f-N_c)$ with $N_f-N_c+1$ quarks
and scale $\Lambda_L^{2(N_f-N_c)-1} = {\det M_0 \over \Lambda^{N_c-1}}
\Lambda^{3(N_f-N_c)-N_f}$ (the factor $\Lambda^{N_c-1}$ in the
denominator is necessary to relate $M_0$ with dimension two, as in the
UV fixed point of the original theory, to that of dimension one
relevant at the UV fixed point of the dual theory).  At lower energies
this theory confines and we should use its gauge invariant observables
\eqn\defin{\eqalign{
N^{\tilde a}_a &=\tilde q^{\tilde a} q_a \cr
b^a&=q^{N_f-N_c} \cr
\tilde b_{\tilde a}&=\tilde q^{N_f-N_c} \cr}}
with $a,\tilde a = 1,..., N_f-N_c+1$.  Their superpotential is
\eqn\lowensp{\Tr \hat M N + {1 \over \Lambda^{2N_f-4N_c+1} \det M_0}
(\tilde b N b - \det N)}
where the first term is \effsup\ expressed in these variables and the
second term is generated dynamically \nati.  Since we now have a low
energy weakly coupled description without any gauge fields it is
straightforward to determine the flat directions.  The superpotential
\lowensp\ is stationary whenever
\eqn\finalfla{\eqalign{
&N=0 \cr
&B \tilde B= -{b\tilde b \over  \Lambda^{2N_f-4N_c+1}} =
\det M_0 \hat M . \cr}}
The relation between $B \tilde B$ and $b\tilde b$ was determined on
dimensional grounds and the numerical coefficient was fixed such that we
recover the known results.

These are precisely the flat directions we found in the electric theory
\flatmbtb\ but now they have a different interpretation.  The electric
gauge group was Higgsed by $\vev{Q}\not=0$.  On the other hand, the
magnetic quarks became massive thus making their low energy theory more
strongly coupled and confining.  This is additional evidence for the
interpretation of the dual variables as magnetic monopoles of the
original ones.  When the electric variables are Higgsed the magnetic
variables are confined.

Note that the electric and magnetic theories remain dual on the flat
directions.  When $B=\tilde B=0$ and $M$ has rank $k$ the low energy
electric theory is an $SU(N_c-k)$ gauge symmetry with $N_f-k$ flavors
while the low energy magnetic theory is its dual which is based on
$SU(N_f-N_c)$ with $N_f-k$ flavors.

\bigskip

\noindent
{\it Mass terms in the magnetic theory}

We now consider the consequences in the magnetic description of mass
terms.  We should add to the superpotential \effsup\ the mass term $\Tr
mM$.  For simplicity let the only non-zero entry of the matrix $m$ be
$m^{\tilde N_f} _{N_f}$ which we will denote by $m$.  Then, we should
study
\eqn\massdefd{M^i_{\tilde i} q_i \tilde q^{\tilde i} +
m M^{N_f}_{\tilde N_f} .}
The equations of motion of $ M^{N_f}_{\tilde N_f}$, $ M^{i}_{\tilde
N_f}$ and $ M^{N_f}_{\tilde i}$ lead to
\eqn\massdeme{\eqalign{
q_{N_f} \tilde q^{\tilde N_f} &= -m  \cr
q_{i} \tilde q^{\tilde N_f} &= 0  \cr
q_{N_f} \tilde q^{\tilde i} &= 0 , \cr}}
(color indices are summed) which show that the gauge group is broken to
$SU(N_f-N_c-1)$ with $N_f-1$ quarks left massless.  The equations of
motion of the massive quarks lead to
\eqn\massdeqe{\eqalign{
M^{N_f}_{\tilde N_f}&=0 \cr
M^{i}_{\tilde N_f}&=0 \cr
M^{N_f}_{\tilde i}&=0 .\cr}}
The low energy superpotential is
\eqn\lowensuph{\hat M^i_{\tilde i} \hat q_i \hat {\tilde q}{}^{\tilde i}
\qquad i,\tilde i = 1,...,N_f-1}
where $\hat M$, $\hat q$ and $\hat{ \tilde q}$ are the light fields with
$N_f-N_c-1$ colors and $N_f-1$ flavors.

In the electric variables the mass term left $N_c$ unchanged and
reduced $N_f$ by one.  Here we found the dual of this result -- both
$N_f$ and $N_f-N_c$ are reduced by one.

The mass term makes the electric variables more strongly coupled.  It is
a relevant operator sending the theory to a more strongly coupled fixed
point.  In the magnetic variables, the gauge group is Higgsed and the
theory becomes more weakly coupled.

Our discussion of the mass term is incomplete for $N_f=N_c+2$ where
the mass term triggers complete breaking of the gauge group.  Then,
the low energy theory should also include instanton contributions in
the broken group.  In all other cases, these are negligible compared
with instantons in the unbroken part of the group and should not be
included.  The relevant instanton calculation is very similar to that
of \ads.  In both cases we study a completely broken $SU(2)$ gauge
theory.  The only difference is that now we have more fermion zero
modes in $q$ and $\tilde q$.  These can be lifted with the interaction
\effsup\ leading to a coefficient proportional to $\det \hat M \over
\Lambda^{N_c+1}$ (again, the power of $\Lambda$ relates $M$ to its
natural normalization at the UV fixed point of the magnetic theory).
Another important factor arises {}from the running of the $SU(2)$
coupling constant, which stops at the scale $\vev{q^{N_f}}=i\sqrt m$
where the symmetry is broken.  Combining all these with the instanton
factor $\Lambda^{4-N_c}$ we find the generated interaction $- {1 \over
\Lambda^{2N_c-2} m} \det \hat M$ and the low energy superpotential
\eqn\lowensup{\hat M^i_{\tilde i} \hat q_i \hat {\tilde q}^{\tilde i}
-{1 \over\Lambda^{2N_c-2} m} \det \hat M.}
Expressing \lowensup\ in terms of $B_i=\Lambda_L^{N_c-1} \sqrt m\hat
q_i$ and $\tilde B^ {\tilde i} =\Lambda_L^{N_c-1} \sqrt m\hat {\tilde q}
{}^{\tilde i}$ (where the powers of $\Lambda_L$ were inserted on
dimensional grounds and the numerical coefficients are as in
\finalfla), it becomes
\eqn\lowensupa{{1 \over \Lambda_L^{2N_c-1}} (\hat M^i_{\tilde i} B_i
\tilde B^{\tilde i} - \det \hat M),}
where $\Lambda_L^{2N_c-1}=m\Lambda^{2N_c-2}$ is the scale of the low
energy theory.  This is the correct superpotential for the low energy
theory with $N_c+1$ flavors \nati.

In \nati\ the superpotential \lowensupa\ was derived by imposing
consistency conditions.  It represents a strong coupling effect which
cannot be understood in weak coupling language.  This fact is
reflected by the positive power of $\Lambda_L$ in the denominator.
Here, we derived this term in a totally weak coupling framework.  The
first term was there at tree level and the second term came {}from
instantons in the broken magnetic group.

One of the consequences of the superpotential \lowensupa\ is the
existence of massless mesons and baryons at the origin of the moduli
space.  Here we get a new perspective on this phenomenon.  The massless
baryons can be interpreted as the dual quarks -- the magnetic monopoles.
It has already been suggested, at least for large $N_c$, that baryons
can be thought of as solitons in the pion Lagrangian
\ref\barsol{T.H.R. Skyrme, Proc.Roy.Soc. {\bf A260} (1961) 127; E.
Witten, \np {160}{1979}{57}; \np{223}{1983}{422}; \np
{223}{1983}{433}}.
Here we see an explicit realization of a related idea -- the baryons
are magnetic monopoles of the elementary quarks and gluons.

An alternate procedure for analyzing the theory with a mass term is to
explore the region in field space with generic $M$ with $B=\tilde B=0$.
Then, all the dual quarks are massive and the low energy magnetic gauge
group leads to gluino condensation.  Working out the $M$ dependence of
the low energy gauge theory, we find that a superpotential proportional
to
\eqn\weffm{{(\det M)^{1 \over N_f-N_c}\over
\Lambda^{3N_c-N_f \over N_f-N_c}}}
is generated.  Adding to it the mass term, $\Tr mM$ we easily find the
expectation values \mexpmass.

\newsec{$SO(N_c)$ theories}

\subsec{Duality in $SO(N_c)$ theories}

In this section we consider $SO(N_c)$ theories with $N_f$ quarks in the
vector representation $Q^i$ ($i=1,...,N_f$ is the flavor index).  One
motivation for studying these theories is that they include theories
exhibiting ordinary duality.  This will enable us to relate our duality
to the better understood duality in Abelian theories.  Another reason
for the interest in these theories is that unlike supersymmetric QCD,
here the quarks are not in a faithful representation of the center of
the gauge group.  Therefore, there is a clear distinction between the
Higgs phase and the confining phase, making the fact that the duality
interchanges them clear.

The global continuous symmetry  of the theory is
\eqn\soglob{SU(N_f)\times U(1)_R}
and the quarks $Q$ transform as
\eqn\qtran{Q \qquad \qquad (N_f,{N_f-(N_c-2) \over N_f}).}
The theory is also invariant under the $Z_{2N_f}$ discrete symmetry
generated by
\eqn\ztnf{Q \rightarrow e^{2\pi i \over 2N_f} Q}
(for $N_c=3$ the symmetry is $Z_{4N_f}$).  The even elements in this
group are also in $SU(N_f)$ so the group is $SU(N_f) \times
Z_{2N_f}/Z_{N_f}$.  We will also be interested in the discrete $Z_2$
``charge conjugation'' symmetry generated by $\CC$ which acts on the
$SO(N_c)$ gauge group like the parity in $O(N_c)$.

The theory is asymptotically free for $N_f < 3(N_c-2)$.  For $N_c
\not=4$ the theory is characterized by a scale $\Lambda$ and for $N_c=4$
there are two independent scales $\Lambda_{1,2}$.  For simplicity, here
we will limit ourselves to the case $\Lambda_1=\Lambda_2$.  Several
special cases of these models have already been studied: $(N_c\ge 6,
N_f=1)$ \ads, $(N_c=3, N_f=1)$ \swi, $(N_c=3, N_f=2)$ \intse, $(N_c=4,
N_f=1)$ \ils, $(N_c=4, N_f=2)$ \intse\ (in the last two examples only
the case of generic $\Lambda_1 \not=\Lambda_2$ was considered) and the
case of arbitrary $(N_c,N_f)$ with $N_f \le N_c-2$ will be discussed in
\ref\isnew{K. Intriligator and N. Seiberg, to appear}.
The picture which emerges {}from these studies is the following.  For
$N_f \le N_c-5$ the gauge group is broken to $SO(N_c-N_f)$ at the
generic point on the moduli space.  In the quantum theory this group
confines and generates a superpotential.  The resulting theory does not
have a ground state.  For $N_f=N_c-4$ or $N_f=N_c-3$ the low energy
theory has two branches.  In one of them the quantum dynamics of the
classically unbroken $SO(N_c-N_f)$ generates a superpotential and in the
other it does not.  For $N_f=N_c-2$ the unbroken group on the moduli
space is $SO(2)$ and there are massless monopole points.

Here we will study the range $N_c-1 \le N_f <3(N_c-2)$.  Repeating the
analysis in the $SU(N_c)$ models (the analogy between the two cases
becomes more clear when the results of the $SO(N_c)$ series are
expressed in terms of $(N_c-2)$) we reach the conclusion that the theory
at the origin is in a non-Abelian Coulomb phase.

The gauge invariant operators which we will be interested in are
\eqn\sonop{\eqalign{
M^{\{ ij\} }&=Q^iQ^j \cr
B^{[i_i,...,i_{N_c}]}&=Q^{i_1}...Q^{i_{N_c}} \cr
b^{[i_i,...,i_{N_c-4}]}&=W_\alpha ^2 Q^{i_1}...Q^{i_{N_c-4}} \cr
\CW_\alpha^{[i_i,...,i_{N_c-2}]}&=W_\alpha Q^{i_1}...Q^{i_{N_c-2}} \cr
}}
where the gauge indices have been suppressed and $W_\alpha$ is the
gauge field strength.

As in the $SU(N_c)$ theories, the interacting non-Abelian Coulomb phase
cannot extend for low values of $N_f$.  We assume that it exists only
for $3(N_c-2)/2<N_f <3(N_c-2)$ where the dimension of the operator $M$
satisfies the unitarity bound.

For $3(N_c-2)/2<N_f < 3(N_c-2)$ a magnetic description of the IR theory is
also possible.  This description can be extended to $N_c-2\le N_f \le
3(N_c-2)/2$ where the magnetic theory is free in the IR and the theory
is in a free non-Abelian magnetic phase.  The magnetic theory is
based on the gauge group $SO(N_f-(N_c-2) +2)$ and it has $N_f$ quarks,
$q$, in the vector representation of the gauge group and the gauge
invariant meson field $M$.  We also assume that under the global
$SU(N_f) \times U(1)_R$ the fields transform as
\eqn\duasot{\eqalign{
q &\qquad (\overline N_f, {N_c-2 \over N_f}) \cr
M &\qquad (\half N_f(N_f+1), 2{(N_f-N_c+2)\over N_f}) . \cr}}
Note that as in the $SU(N_c)$ theories, the $SU(N_f)$ transformation
laws of $q$ are dual to those of $Q$.  The assignment of the $R$ charge
is anomaly free.  As in the $SU(N_c)$ theories, we add an invariant
superpotential
\eqn\macsosu{W=M^{\{ ij\} }q_iq_j.}
(For $N_f=N_c-1, N_c-2$ and for $N_c=3$ the magnetic theory is more
complicated.)

We check the consistency of our ansatz by examining the 'tHooft anomaly
conditions.  We find for both theories the same anomalies:
\eqn\thooftso{\eqalign{
SU(N_f)^3 \qquad &N_c d^{(3)}(N_f) \cr
SU(N_f)^2U(1)_R \qquad & -{N_c(N_c-2) \over N_f}d^{(2)}(N_f) \cr
U(1)_R \qquad &-{N_c(N_c-3) \over 2} \cr
U(1)_R^3 \qquad &{N_c \over 2}(N_c-1-2{(N_c-2)^3 \over N_f^2}) . \cr}}

It is important that the electric theory and its magnetic dual have the
same gauge invariant operators.  It is easy to check that the operator
$B$ in \sonop\ has the same quantum numbers as the operator $b$ of the
dual, magnetic theory.  Similarly, the operator $b$ of the electric
theory has the same quantum numbers as the operator $B$ of the magnetic
theory.  Therefore, it is natural to identify these pairs.  Furthermore,
the ``field strength'' $\CW_\alpha^{[i_1...i_{N_c-2}]}$ in the two
theories have the same quantum numbers and therefore can be identified
(more precisely be proportional to each other).  $\CW$ can be
interpreted as the $SO(N_c)$ gauge invariant description of the light
photon along the flat direction where $SO(N_c)$ is broken to $SO(2)$.
The identification between these operators in the two theories is the
non-Abelian generalization of the duality studied in \swi.  There, for
$N_c=3$, $N_f=1$, the field strength of the low energy photon on the
moduli space is $\Tr W_\alpha Q$.  The dual, magnetic theory includes
the monopole field $q$ and the magnetic photon, whose field strength
is indeed proportional to $\Tr W_\alpha Q$.

In checking the identification of the operators it is also important to
examine their transformation laws under the discrete symmetries.  The
discrete charge conjugation $\CC$ is mapped to $\CC$ of the magnetic
theory.  However, the $Z_{2N_f}$ symmetry is more subtle.  This symmetry
in the magnetic theory is generated by $q \rightarrow \CC e^{-{2\pi i
\over 2N_f}}q$.  The factor of $\CC$ in the generator ensures that the
operators in the two theories are mapped as we said above.

The analysis of the flat directions and the effect of mass terms is
similar to the $SU(N_c)$ theories.  The mass terms make the electric
theory more strongly coupled.  They induce gauge symmetry breaking in
the magnetic description.  The flat directions break the electric gauge
symmetry and make the magnetic theory more strongly coupled.  As in the
$SU(N_c)$ series, these analyses become more subtle when we approach the
boundaries of the phase: $N_f=(N_c-2)+1$ and $N_c=3$.

To demonstrate the physics of the flat directions and mass perturbations
we consider the special case of $N_f=N_c$.  The magnetic gauge theory is
$SO(4)$ with $N_f$ flavors.  Consider the region in field space with
generic $M$ in the magnetic theory.  All the magnetic quarks are massive
and the low energy $SO(4) \cong SU(2)_1 \times SU(2)_2$ pure gauge
theory has a scale $\Lambda_L^6 \sim \det M \Lambda^{6-2N_f}$ (the power
of $\Lambda$ is determined by the one loop beta function and the fact
that we use the dimension of $M$ as in the electric UV fixed point
(two)).  This theory becomes strongly coupled and has four vacua with
$\vev{(\lambda\lambda)_1} \sim \pm \Lambda_L^3$ and
$\vev{(\lambda\lambda)_2} \sim \pm \Lambda_L^3$ corresponding to the
(magnetic) gluino condensation in the two $SU(2)$ subgroups of the
magnetic theory. This leads to a superpotential
$W=2\vev{(\lambda\lambda)_1} + 2\vev{(\lambda\lambda)_2} $.  It has two
branches.  In one of them $W=0$ and in the other $W$ is proportional to
$ \pm \Lambda^{3-N_f}(\det M)^{1
\over 2}$.

Consider now the effect of a mass term $\Tr mM$, which induces
confinement of the electric variables.  The branch with the vanishing
superpotential does not lead to a supersymmetric ground state.  The
other branch leads to $N_c-2$ vacua $\vev{M} \sim \Lambda^{2N_c-6 \over
N_c-2} (\det m)^{1 \over N_c-2} {1 \over m}$.  To relate this phase to
the $m=0$ theory note that for small $m$ at least some of the components
of $M$ are small.  Then a more appropriate description, which we will
study in detail below, is in terms of expectation values of $q$.  Hence,
the magnetic theory is Higgsed corresponding to confinement of the
electric variables.

For the flat directions we should pick the branch with
$\vev{(\lambda\lambda)_1} =-\vev{(\lambda\lambda)_2}$, where the
superpotential vanishes.  Note that every value of $M$ leads to two
ground states differing by the sign of $\vev{(\lambda\lambda)_1 -
(\lambda\lambda)_2}$.  In the electric variables these vacua are labeled
by $M$ and the baryon field $B=\det Q$ which is constrained by $\det M =
B^2$.  The two vacua for a given $M$ differ by the sign of $B$.  Above
we suggested that the operator $B$ be identified with the operator $b$
of the magnetic theory.  In this case this leads to the identification
of $B$ with $(W_\alpha)^2_1 -(W_\alpha)^2_2$ of the magnetic theory.  We
see that the expectation values of these operators are compatible with
the suggested identification.

It has already been noticed in \intse\ that in different phases the
effective superpotential is different.  Here we see the different
branches arising as different values of the gluino condensates in the
magnetic group.

\subsec{$N_f=N_c-1, N_c-2, N_c-3$}

We start by discussing the mass perturbation which should decouple a
single quark in the $N_f=N_c$ theory.  This is analogous to decoupling a
quark in the $N_f=N_c+2$ theory with $SU(N_c)$ gauge group.  The
magnetic gauge group is $SO(4)$.  The mass term induces a gauge symmetry
breaking to $SO(3)$.  When the gauge group is larger we do not have to
consider the effect of instantons in the broken part of the group.
However, in this case, $SO(4)$ instantons which are not in $SO(3)$ are
well defined and their effect should be represented in the low energy
theory as a local operator in the Lagrangian.  The calculation of this
term is similar to the calculation in the analogous $SU(N_c)$ theory and
was described above.  The induced term is proportional to
\eqn\indsot{-{1 \over \Lambda_L^{2(N_c-1)-3}} \det \hat M}
where $\hat M$ is the matrix of light mesons and $\Lambda_L$ is the
scale of the low energy electric theory.

The lesson {}from this calculation is that the dual of an $SO(N_c)$ theory
with $N_f=N_c-1$ is an $SO(3)$ gauge theory with the superpotential
\eqn\sothdusu{M^{ij}q_i q_j - {1 \over \Lambda^{2(N_c-2)-1}} \det  M}
(we absorb a constant in a redefinition of $\Lambda$).
It is easy to check that the second term is invariant under all the
symmetries of the electric theory.  In fact, if it is left out the
$SO(3)$ theory has a $Z_{4N_f}$ symmetry ($q \rightarrow e^{-{2\pi i
\over 4N_f}}q$) that the electric theory does not have.

Starting {}from the superpotential \sothdusu\ for $N_f=N_c-1$ we can
decouple another quark and flow to the theory with $N_c-2$ flavors.
We add $m M^{N_f N_f}$ to \sothdusu\ and integrate out the massive
fields.  As above, we denote by $\hat M$ the components of $M^{ij}$
with $i,j \not= N_f$ and for simplicity, we set $\Lambda=1$.  The
equation of motion of $M^{N_fN_f}$ leads to $q_{N_f}q_{N_f}= \det \hat
M -m$.  For generic $\det \hat M$ it leads to a non-zero expectation
value for $q_{N_f}$ which breaks the $SO(3)$ gauge symmetry to
$SO(2)$.  The low energy theory around that point is an $SO(2)$
theory with $N_c-2$ quarks in two dimensional representations $\hat
q_i$ and the light mesons $\hat M$ coupled by a superpotential
\eqn\effsupsotw{\hat M^{ij} \hat q_i \hat q_j.}

This analysis is not valid when $\det \hat M$ is too large.  Then, the
first term in \sothdusu\ leads to a mass term to the first $N_f-1$
quarks and the low energy theory is an $SO(3)$ theory with one triplet
$q_{N_f}$ with the superpotential
\eqn\lowenwsot{M^{N_fN_f}(q_{N_f}q_{N_f}-\det \hat M +m)}
its scale is $\Lambda_L^4 \sim (\det \hat M)^2$.  This theory is
strongly coupled.  Without the superpotential \lowenwsot\ it is the
theory studied in \swi.  The coupling to $M^{N_fN_f}$ acts like a mass
term for $q^{N_f}$ which locks the theory at two ground states with
$\vev{q_{N_f}q_{N_f}} = \pm \Lambda_L^2\sim \pm \det \hat M$ leading to
the superpotential
\eqn\lowenwsot{M^{N_fN_f}(\pm \det \hat M -\det \hat M +m)}
(the numerical coefficients were arranged to agree with the
semiclassical answer at large $M$).  The ground state of the $SO(3)$
dynamics with $\vev{q_{N_f}q_{N_f}}\sim \det \hat M$ does not satisfy
the $M^{N_fN_f}$ equation of motion and therefore does not lead to a
ground state of the complete theory.  The other solution leads to a
ground state with $\det \hat M =m/2$ and $M^{N_fN_f}=0$.  In order to
find the effective superpotential around that point we should remember
that the low energy superpotential includes a light monopole (a doublet
of $SO(2)$) $E$ and the field $u=q_{N_f}q_{N_f}$
\eqn\anothereffw{M^{N_fN_f}(u -\det \hat M +m) + (u+ \det
\hat M)E^2.}
Integrating out $M^{N_fN_f}$ and $u$ we find our final answer near
that point
\eqn\fineffwso{(2\det \hat M -m)E^2 .}

We conclude that the theory with a mass term has a moduli space labeled
by $\hat M$ which is in the Coulomb phase.  Since there are massless
magnetic monopoles we can refer to this phase as a ``free Abelian
magnetic phase.''  There are two singular submanifolds $\det \hat M = 0,
m \Lambda^{2N_c-5}/2$ (we have restored the power of $\Lambda$ on
dimensional grounds) with massless charged fields.  Both of these
submanifolds are strongly coupled in the electric variables.  However,
the submanifold with $\det \hat M=0$ is weakly coupled in the magnetic
variables and the light fields $\hat q$ are the magnetic quarks while
the submanifold with $\det \hat M = m \Lambda^{2N_c-5}/2$ is also
strongly coupled in the magnetic quarks.  The massless $\hat q$ are
clearly monopoles in the electric variables.  Here we identify them as
components of the magnetic quarks.  This is further evidence for our
interpretation of the dual quarks as magnetic monopoles.

We can use the result of this discussion as the solution of the theory
with $N_f=N_c-2$.  Its moduli space is in the Coulomb phase and is
labeled by a symmetric $N_f$ dimensional matrix $M$.  The
singularities are at $\det M= 0, \Lambda^{2(N_c-2)}/2$ with a single
massless monopole $E$ at $\det M= \Lambda^{2(N_c-2)}/2$ and $N_f$
massless monopoles $q_i$ coupled to $M$ by the superpotential
$M^{ij}q_iq_j$.  The theory around $M=0$ is exactly what we would have
found by simply applying our duality rules to the electric theory.

We can also use this solution to continue to flow down to $N_c-3$
flavors by turning on a mass term $m M^{N_fN_f}$.  The moduli space is
again labeled by $\hat M$ (the components of $M$ which do not involve
the massive quark).  Integrating out all the fields except $\hat M$, we
find two branches.  One of them arises {}from the single monopole point.
The monopole condenses and lifts the photon and a superpotential
proportional to $1 \over \det \hat M$ is generated.  The other branch
arises {}from the other monopole point.  Using \effsupsotw, we find that
$q_{N_f}$ condenses and lifts the photon.  The massless fields are $\hat
M$ and $N_f-1$ of the magnetic quarks $\hat q$ without color indices
coupled by the superpotential $\hat M^{ij}\hat q_i \hat q_j$.

We thus find the solution of the theory with $N_f=N_c-3$.  It has one
branch with a superpotential proportional to $1 \over \det M$.  The
light fields in the other branch are $M^{ij}$ and $q_i$ coupled by
$M^{ij} q_i q_j$.  The massless fields $q_i$ can be interpreted as
solitons of the elementary quarks.  Alternatively, they can be thought
of as the gauge invariant ``exotic'' $q \sim b= W_\alpha^2 Q^{N_c-4}$ of
equation \sonop\ (for $N_c=4$ this is a glueball).  As with the baryons
in the $SU(N_c)$ series with $N_f=N_c+1$, we see that such bound states
can be interpreted as solitons of the elementary quarks.

More aspects of these $SO(N_c)$ theories will be discussed in \isnew.

\bigskip

\centerline{{\bf Acknowledgments}}

We would like to thank M.R. Plesser, P. Pouliot, S. Shenker, E. Witten
and especially K. Intriligator for useful discussions and comments.
This work was supported in part by DOE grant
\#DE-FG05-90ER40559.

\listrefs

\end